\documentclass{ws-ijmpa}

\begin{document}

\markboth{R. Maciu{\l}a, A. Szczurek and G. {\'S}lipek}
{Production of D and B mesons and their semileptonic decays}

\catchline{}{}{}{}{}
%
%%%%%%%%%%%%%%%%%%%%%%%%%%%%%%%%%%%%%%%%%%%%%%%%%%%%%%%%%%%%%%%%%%%%

\title{PRODUCTION OF D AND B MESONS AND THEIR SEMILEPTONIC DECAYS}

\author{RAFA{\L} MACIU{\L}A}
\address{Institute of Nuclear Physics PAN, PL-31-342 Cracow, Poland\\
rafal.maciula@ifj.edu.pl}

\author{ANTONI SZCZUREK}
\address{University of Rzesz\'ow, PL-35-959 Rzesz\'ow, Poland\\
Institute of Nuclear Physics PAN, PL-31-342 Cracow, Poland\\
antoni.szczurek@ifj.edu.pl}

\author{GABRIELA {\'S}LIPEK}
\address{Institute of Nuclear Physics PAN, PL-31-342 Cracow, Poland\\
gabriela.slipek@ifj.edu.pl}

\maketitle

%\begin{history}
%\received{Day Month Year}
%\revised{Day Month Year}
%\end{history}

\begin{abstract}
We have calculated the kinematical correlations between charged leptons
from semileptonic decays of open charm/bottom as well as leptons
produced in the Drell-Yan mechanism in proton-proton collisions at BNL
RHIC. Presented production rates for open charm and bottom were
estimated by a detailed analysis of so-called nonphotonic electrons. The
distributions for charm and bottom quarks pairs are calculated in the
$k_{t}$-factorization approach with the help of the Kwieci{\'n}ski
unintegrated parton distributions. The hadronization of heavy quarks is
done by means of Peterson et al. fragmentation function. The
semileptonic decay functions are found by fitting recent semileptonic
data obtained by the CLEO and BABAR. We get good description of
recent PHENIX data.

\end{abstract}

\ccode{PACS numbers: 12.38.-t, 12.38.Cy, 14.65.Dw}

\section{Introduction}	

    Recently the PHENIX collaboration
has measured dilepton invariant mass spectrum from $0$ to $8$ GeV in proton-proton collisions at $\sqrt{s}=200$ GeV\cite{PHENIX}.
It is believed that the main contribution
to the dielectron continuum comes from so-called nonphotonic electrons which are produced in semileptonic decays of open charm and bottom mesons. To date, productions of charm and bottom was usually studied by standard measurements\cite{electrons} and pQCD calculations\cite{theory} of single leptons inclusive distributions. Such predictions give rather good description of the experimental data, however the theoretical uncertainties are quite large what makes the situation somewhat clouded and prevents definite conclusions.

%=======================

 Better statistics at present colliders gives a new possibility to study
 not only inclusive distributions but also correlations between outgoing
 particles. Kinematical correlations constitute an alternative method to
 study open charm and bottom production. It gives also a great possibility to separate charm and bottom contributions what has a crucial meaning for understanding the character of heavy quarks interactions with the matter created in high energy nuclear collisions\cite{mischke}.   

%===============
 The process of nonphotonic electron production has three subsequent stages.
The whole procedure can be written in the following schematic way:
\begin{equation}
\frac{d \sigma^e}{d y d^2 p} =
\frac{d \sigma^Q}{d y d^2 p} \otimes
D_{Q \to h} \otimes
f_{h \to e} \; ,
\label{whole_procedure}
\end{equation}
where the symbol $\otimes$ denotes a generic convolution.
The first term is responsible for production
of heavy quarks/antiquarks. Next step is the process of formation of
heavy mesons and the last ingredient describe semileptonic decays of 
heavy mesons to electrons/positrons.
%===========================================================================

\vskip-6.5mm
\section{Formalism}
 The inclusive production of heavy quark/antiquark pairs can be calculated
in the framework of the $k_t$-factorization\cite{CCH91}.
In this approach transverse momenta of initial partons are included and
the emission of gluons is encoded in 
so-called unintegrated gluon (parton) distributions.
In the leading-order approximation within the $k_t$-factorization approach
the differential cross section for the $Q \bar Q$ or Drell-Yan process can be written as:
\begin{eqnarray}
\frac{d \sigma}{d y_1 d p_{1t} d y_{2} d p_{2t} d \phi} =
\sum_{i,j} \; \int \frac{d^2 \kappa_{1,t}}{\pi} \frac{d^2 \kappa_{2,t}}{\pi}
\frac{1}{16 \pi^2 (x_1 x_2 s)^2} \; \overline{ | {\cal M}_{ij} |^2}\\
\nonumber 
\delta^{2} \left( \vec{\kappa}_{1,t} + \vec{\kappa}_{2,t} 
                 - \vec{p}_{1,t} - \vec{p}_{2,t} \right) \;
{\cal F}_i(x_1,\kappa_{1,t}^2) \; {\cal F}_j(x_2,\kappa_{2,t}^2) \; , \nonumber \,\,
\end{eqnarray}
where ${\cal F}_i(x_1,\kappa_{1,t}^2)$ and ${\cal F}_j(x_2,\kappa_{2,t}^2)$
are the unintegrated gluon (parton) distribution functions (UPDFs). 

%===========================================================================
 
There are two types of the LO $2 \to 2$ subprocesses which contribute
to heavy quarks production, $gg \to Q \bar Q$ and $q \bar q \to Q \bar Q$.
The first mechanism dominates at large energies and the second one
near the threshold. At relatively low RHIC energies rather intermediate $x$-values
become relevant so the Kwiecinski UPDFs seem applicable in this case\cite{Kwiecinski}.

%===========================================================================
The hadronization of heavy quarks is usually done
with the help of fragmentation functions. The inclusive distributions of hadrons
can be obtained through a convolution of inclusive distributions
of heavy quarks/antiquarks and Q $\to$ h fragmentation functions.
The Peterson fragmentation functions are often used in this context\cite{Peterson}.

%===============================================================================
 Recently the CLEO and BABAR 
collaborations have measured very precisely
the spectrum of electrons/positrons coming from
the weak decays of $D$ and $B$ mesons, respectively\cite{CLEO}.
These functions can in principle be calculated.
This introduces, however, some model 
uncertainties and requires inclusion of all final 
state channels explicitly. An alternative is to use 
proper experimental input which after renormalizing 
to experimental branching fractions
can be use to generate electrons/positrons 
in a Monte Carlo approach.

%===============================================================================
 The $k_t$-factorization method is very useful to study correlations between produced leptons because it can reproduce distributions in kinematical variables related to the transverse momentum of initial partons. In this part of our calculations we take under consideration not only leptons from open charm/bottom decays but also leptons produced in Drell-Yan proccess, as well as leptons coming from elastic and inelastic processes initiated by photon-photon fusion. In the case of elastic reaction we follow exact momentum space calculations with 4-body phase space and for inelastic scattering we have applied unique photon distributions in the nucleon MRST2004\cite{MRST}.

\vspace{-1.5mm}
\section{Results and Conclusions}
\vskip-6.65mm
\begin{figure}[!thb]
\begin{center}
 \includegraphics[width=7.5cm]{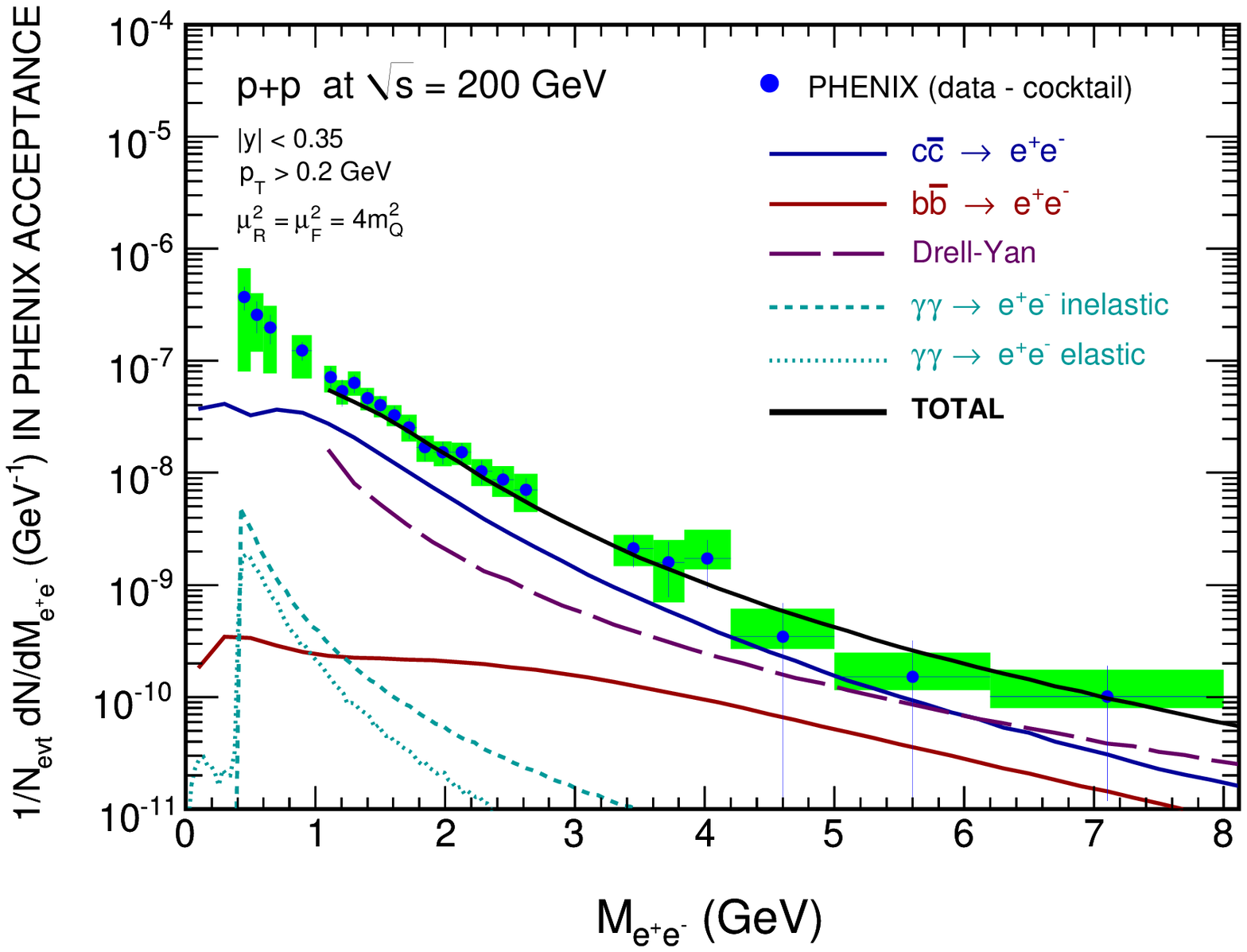}
 \includegraphics[width=5cm]{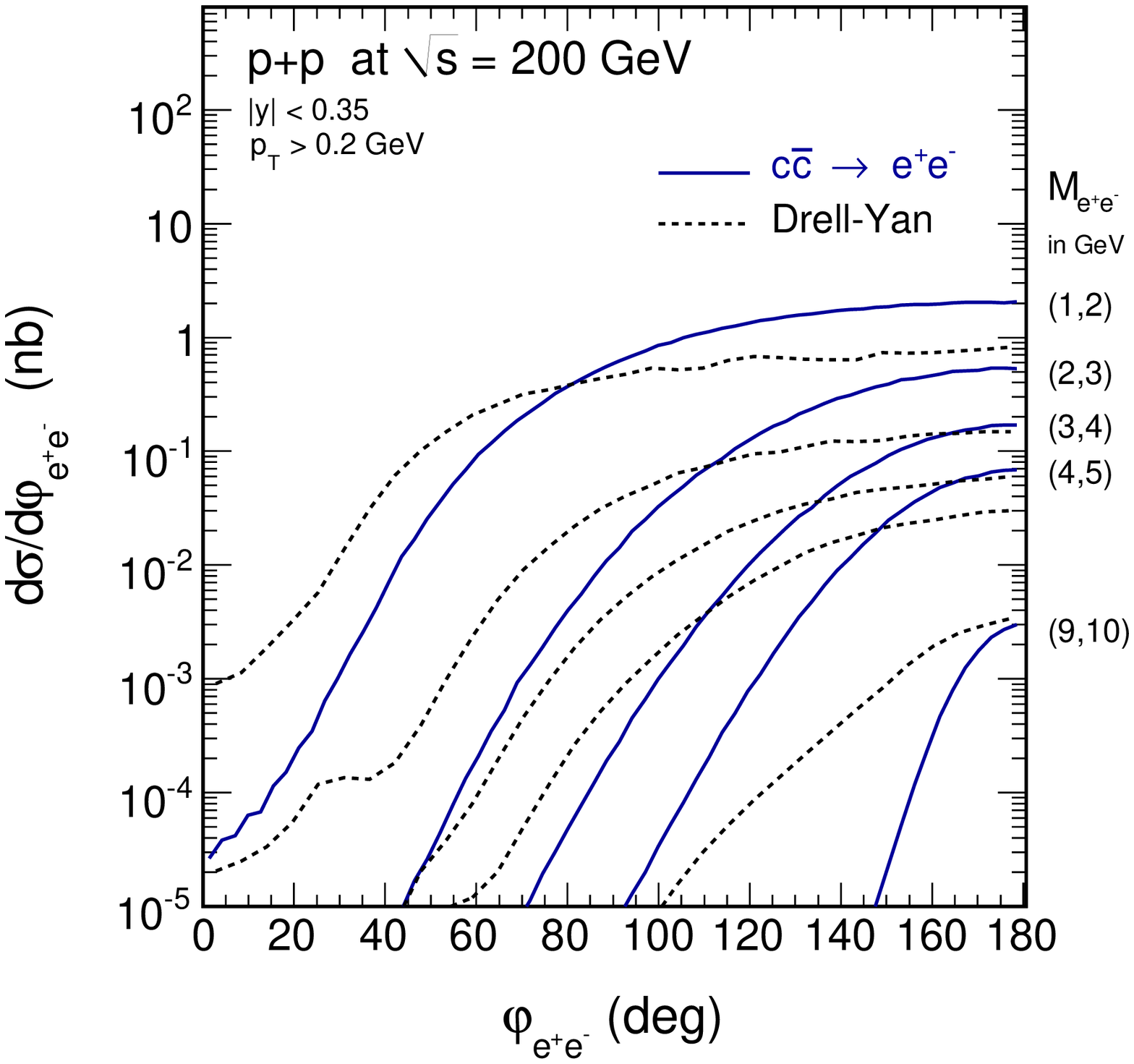}
\vskip-4mm\caption[*]{
Dilepton invariant mass spectrum with PHENIX data and our predictions (left panel) and azimuthal angle correlations between nonphotonic and Drell-Yan leptons in different invariant mass regions (right panel).}
\end{center}
\vskip-0.5mm
\end{figure}
\vskip-6.5mm
We have calculated dilepton mass distribution as well as azimuthal angle
correlations between outgoing leptons. We get a rather good description
of the recent PHENIX data.

\end{document}